# UCN anomalous losses and the UCN capture cross-section on material defects

A. Serebrov, N. Romanenko, O. Zherebtsov,

M. Lasakov, A. Vasiliev, A. Fomin,

I. Krasnoshekova, A. Kharitonov, V. Varlamov

*Petersburg Nuclear Physics Institute, Russian Academy of Sciences, Gatchina, Leningrad District, 188300, Russia*

**Abstract**

Experimental data shows anomalously large Ultra Cold Neutrons (UCN) reflection losses and that the process of UCN reflection is not completely coherent. UCN anomalous losses under reflection cannot be explained in the context of neutron optics calculations. UCN losses by means of incoherent scattering on material defects are considered and cross-section values calculated. The UCN capture cross-section on material defects is enhanced by a factor of $10^4$ due to localization of UCN around defects. This phenomenon can explain anomalous losses of UCN.

**1. Introduction**

Right from the first experiments studying the storage of ultra cold neutrons in material traps [1, 2], observed reflection losses are significantly higher than expected from the neutron optics calculations for perfect surfaces. Later experiments revealed the phenomenon of UCN heating [3] due to hydrogen layers on the reflecting surface. However, the hydrogen concentration on the surface was not high enough to explain the UCN losses, which were at the level of $10^{-3}$. By using materials with low neutron absorption and cooling the surface down to low temperatures, UCN loss factors, $\mu$, at the level of 2 to $3 \cdot 10^{-5}$ down to $2 \cdot 10^{-6}$ have been obtained: Table 1 collects the lowest $\mu$ values measured so far for several materials, Fomblin oil at room temperature, Beryllium, solid Oxygen and low-temperature (frozen) Fomblin oil (LTF).

Table 1. Presently observed lowest UCN losses factors for different materials.

| Fomblin (300 K) | $\mu = (2.2 \pm 0.1) \cdot 10^{-5}$ | [4] |
|---|---|---|
| *Be* (6.5 K) | $\mu_{Be} = 3.2 \cdot 10^{-5}$ | [5] |
| *Be* (10 K) | $\mu_{Be} = (2.8 \pm 0.4) \cdot 10^{-5}$ | [6] |
| $O_2$ (10 K) | $\mu_{O_2} = (6.1 \pm 0.6) \cdot 10^{-6}$ | [6] |
| LTF (110 K) | $\mu = (2.2 \pm 0.2) \cdot 10^{-6}$ | [7] |

The difference between the expected and measured loss factors for Beryllium are particularly large. The interaction cross-sections for neutrons in the velocity range 10 to 12 m/s have been measured and based on these results, the expected loss factor for UCN in wall collisions should be less than $2 \cdot 10^{-6}$. However, experiments show a loss factor of $3 \cdot 10^{-5}$ at temperatures of 6 to 10 K [5, 6].

For solid oxygen the calculated loss factor is one order of magnitude less than measured. However, as there are no experimental cross-sections for inelastic scattering at low neutron energies and low temperatures, no conclusions can be made.

There is no contradiction between the calculated and measured loss factors for Fomblin oil. At room temperature a process of quasi-elastic scattering on the surface waves of liquid has been observed; this explains the loss factor of $2 \cdot 10^{-5}$. This process is suppressed in the case of LTF [7, 8] and the loss factor at a temperature of 110 K, when it is solid, $2.2 \pm 0.2 \cdot 10^{-6}$ [7], is consistent with the measured inelastic scattering cross-section [9].

## 2. Studies of temperature and energy dependence of losses on Beryllium

The temperature and energy dependence of losses of UCN stored in a beryllium-coated trap were studied during the course of preparation work for the neutron lifetime experiment at the "Gravitrap" installation [6, 7].

The measured temperature dependence of the UCN loss factor, $\mu$, during storage in a Beryllium trap are shown in Fig 1. The UCN loss factor depends on the concentration of hydrogen on the surface of the trap and the trap temperature. At low temperatures, when the up-scattering process is suppressed completely, the UCN loss factor is, nevertheless, still about $3 \cdot 10^{-5}$ and exceeds the expected value by more than an order of magnitude: This effect has been called the UCN anomalous loss for beryllium.

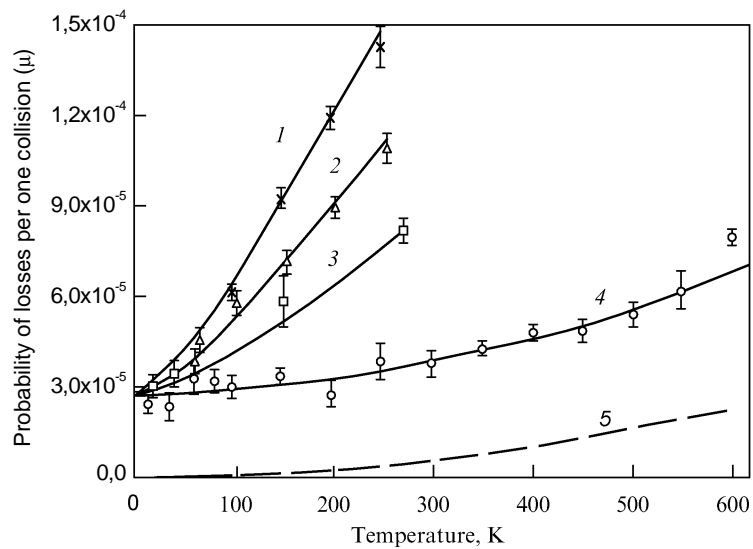

**Figure 1:** Temperature dependence of the UCN loss factor $\mu$ for different beryllium traps:
1 – spherical deposited *Be* trap, not degassed;
2 – cylindrical deposited *Be* trap, degassed (5 hours at 250°C);
3 – whole *Be* trap, degassed (8 hours at 300°C);
4 – spherical deposited *Be* trap degassed
     (28 hours at 350°C with purification of the evaporated *He* and $D_2$);
5 – theoretical temperature dependence calculated in the Debye model.

Fig. 2 shows the measured energy dependence of the probability of losses $\tau_{loss.}^{-1}$ in a *Be* trap at a temperature of 90 K. It should be noted that due to the gravitational field, the relationship between $\tau_{loss.}^{-1}(E)$ and $\mu_{exp}(E)$ is not direct but it can be calculated easily. The solid curves in Fig. 2 represent the calculated energy dependence of the losses; these curves have been normalized to the reflectivity measured at the lowest energy in both cases.

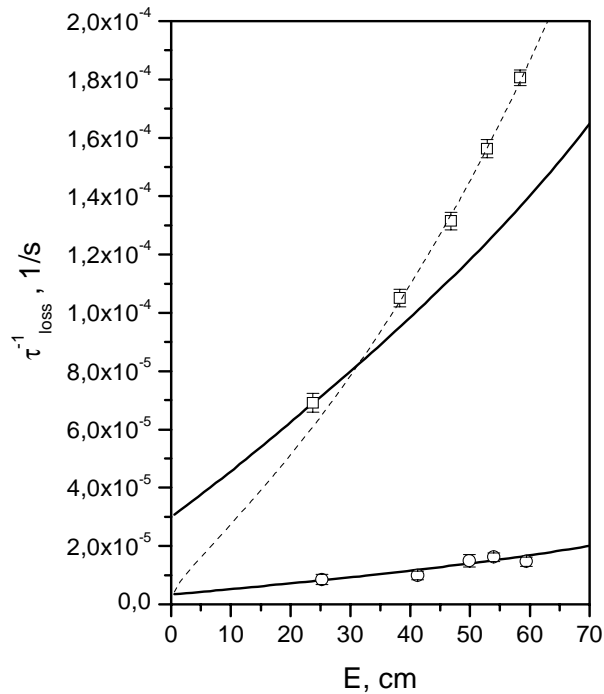

**Figure 2:** The neutron energy dependence of the probability of UCN losses for Beryllium trap at 90 K ( ) and for the trap with low temperature Fomblin oil at 110 K (O). The solid lines are fits to the expected theoretical dependence of normal UCN losses.

While there is no contradiction for the LTF coating, we can see that the experimental and theoretical energy dependence in the case of Beryllium are very different. That is, the UCN loss factor for Beryllium cannot be described in the context of neutron optics considerations.

## 3. Studies of UCN transmission by Beryllium samples and Beryllium coatings

Study of the transmission of neutrons in the velocity range 10 to 12 m/s through various Beryllium samples was carried out as part of the work of developing an UCN converter at the WWR-M reactor. Fig. 3 shows the temperature dependence of the macroscopic cross section for the interaction of very cold neutrons with different material samples of Beryllium, extruded, fused and quasi-monocrystal. We can see that at low temperatures, when the up-scattering cross section is suppressed, the macroscopic cross section for elastic scattering on material defects is in the range 1 to 2 cm$^{-1}$. However, using this value in calculations of the UCN loss factor in the context of neutron optics does not explain the anomalous losses.

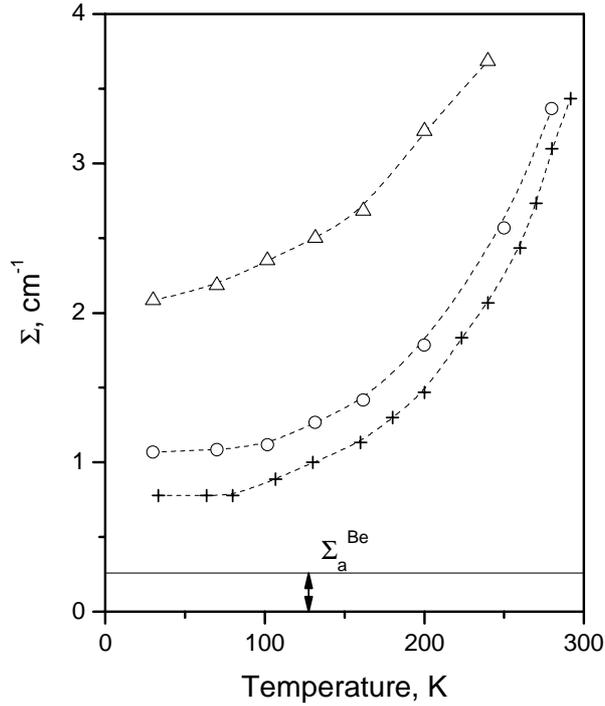

**Figure 3:** The temperature dependence of the macroscopic cross-section ($\sum$) for extruded beryllium ($\Delta$), quasi-monocrystal beryllium (O), fused beryllium (+). The macroscopic capture cross-section for $Be$ nuclei $\sum_a^{Be} = \rho^{Be} \cdot \sigma_a^{Be}$ is shown by a straight line.

Measurement of the UCN transmission through Beryllium coatings sputtered onto aluminum foils (15 μm), copper foils (10 μm) and silicon wafers (350 μm) were carried out at the ILL reactor using PNPI's gravitational spectrometer. The primary purpose of these studies was to examine the quality of sputtered coatings. However, the experimental results can also be used to see if anomalous reflection losses for UCN by Beryllium are accompanied by anomalous transmission values.

The transmission for fifteen different samples was measured for UCN in the energy range 150 to 280 neV, from just higher than the boundary energy of the substrate to just below the boundary energy of beryllium. The results are shown in Fig. 4. The transmission probability for Beryllium coatings with a thickness of 2000 to 6000 Å on Aluminum and Copper foils varied from $2 \cdot 10^{-5}$ to $1 \cdot 10^{-4}$ and have an average value of $4 \cdot 10^{-5}$.

Some of the foils were coated twice on the same side; cleaning the surface with alcohol after the first coating to move dust and then re-coating the damaged locations: This did not help to reduce the transmission.

Foils sputtered on both sides did not transmit any neutrons, which, under the conditions of the experiment, corresponds to a transmission factor of $< 1 \cdot 10^{-8}$.

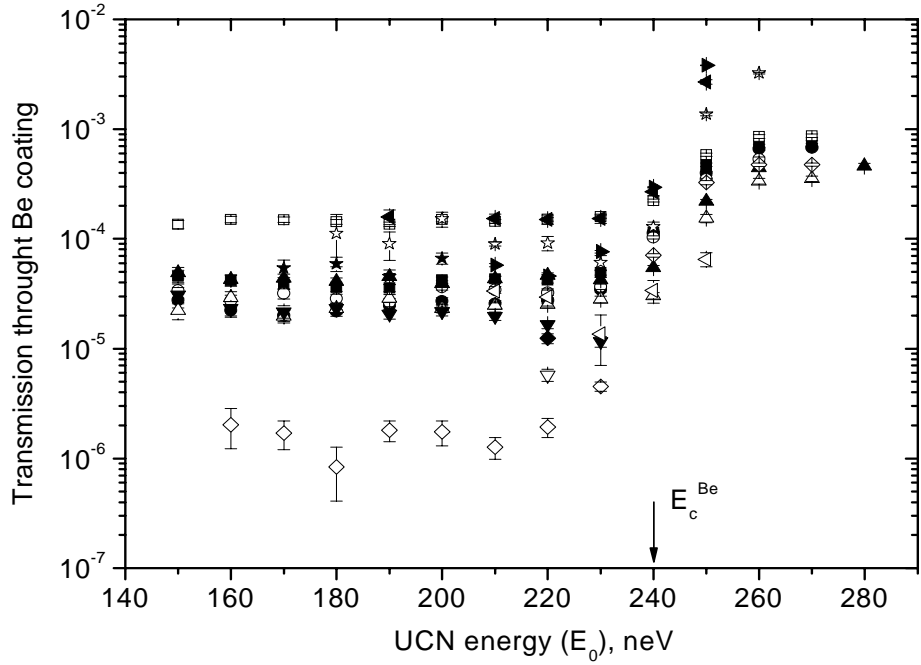

**Figure 4:** The transmission through *Be* foils
■ is *Al* (20 µm) + *Be* (2150 Å) + *Be* (2150 Å)*;
□ is *Al* (20 µm) + *Be*(2150 Å) + *Be*(2150 Å);
● is *Al* (15 µm) + *Be* (2150 Å) + *Be* (2150 Å)*;
○ is *Al* (20 µm) + *Be* (2150 Å) + *Be* (2150 Å)*;
▲ is *Al* (15 µm) + *Be* (3500 Å);
∆ is *Al* (15 µm) + *Be* (6100 Å);
▼ is *Si* wafer 300 µm + *Be* (2150 Å);
∇ is *Si* wafer 300 µm + *Be* (2150 Å);
◆ is *Si* wafer 300 µm + *Be* (2150 Å) + *Be* (2150 Å)*;
◇ is *Si* wafer 300 µm + *Be* (2150 Å) + *Be* (2150 Å)*;
★ is *Si* wafer 300 µm + *Be* (2150 Å);
☆ is *Cu* (10 µm) + *Be* (6100 Å);
◀ is *Cu* (10 µm) + *Be* (3500 Å);
◁ is *Cu* (10 µm) + *Be* (3500 Å);
▶ is *Cu* (10 µm) + *Be* (3500 Å).
* twice coated on the same side with the cleaning of surface with alcohol after the first coating.

$E_c^{Be}$ is the critical energy for UCN reflection on *Be*.

The transmission probability for the sputtered silicon wafers appears to be 3 to 5 times less than for the other samples and for one was $2 \cdot 10^{-6}$. The distinctive feature of the Si-wafers is the polished surface of a single crystal, which allows a higher quality coating to be produced. However, the value of $2 \cdot 10^{-6}$ is apparently only a technological limit and due to dust on the surface; the same light transmission coefficient, 2 to $4 \cdot 10^{-6}$ was obtained for coatings of different metals sputtered under the same conditions over glass substrates.

The results indicate that the average fraction of defects on the surface of Beryllium coated Aluminum and Copper substrates is $4 \cdot 10^{-5}$. However, in the case of reflection of UCN with energy below the critical energy for Aluminum or Copper, such defective surfaces should not be important as neutrons can also be reflected from the substrate. That is, if the surface defects are the cause of the anomalous losses of UCN, it would mean that scattering of UCN by these defects also causes a significant part of the UCN to be absorbed. This, in turn, requires that scattering on a material defect is incoherent; the existence of an incoherent scattering component in the wall collision can be confirmed by observing UCN depolarization during their storage in traps.

## 4. Depolarization of UCN reflected from Beryllium and its possible connection to the phenomenon of anomalous losses

The process of UCN depolarization while stored in a Beryllium trap was first observed about four years ago and studied further in subsequent experiments [10, 11]; this is an experimental demonstration that there is an incoherent component to the scattering of UCN reflecting from matter. The depolarization process is found to be temperature independent and shows that the UCN incoherently scattered into material do not return to the vacuum.

Measured UCN spin-flip probabilities for various Beryllium samples are collected into Table 2. The probability of a spin-flip, $\alpha$, is about 1 to $2 \cdot 10^{-5}$ per collision and very close to the probability of anomalous losses. This suggests there is some connection between the depolarization and the anomalous losses, with both due to an incoherent scattering process on the defects in the beryllium surface.

Table 2. The probability of an UCN spin flip per collision for *Be* samples at room temperature. The measurement uncertainties include only the statistical accuracy of the measurements. The calculations of the number of collisions has an accuracy of between 20 and 30 % and thus contributes correspondingly to the absolute experimental uncertainties.

| Material | $\alpha$ |
|---|---|
| Trap coating (*Be*) I | $(0.72\pm0.07)\cdot10^{-5}$ |
| Trap coating (*Be*) II | $(2.07\pm0.05)\cdot10^{-5}$ |
| *Be* foil 1 | $(1.58\pm0.20)\cdot10^{-5}$ |
| *Be* foil 2 | $(2.17\pm0.21)\cdot10^{-5}$ |
| *Be* coating on copper rings | $(1.15\pm0.09)\cdot10^{-5}$ |
| *Be* coating on *Al* foil | $(1.23\pm0.21)\cdot10^{-5}$ |

Measurement of the ratio of the loss probability to the depolarization probability is of fundamental importance, as it allows necessary information concerning the possible connection between the two phenomena to be obtained. In particular, a simplified model of the phenomenon predicts that when the fractions of UCN incoherently scattered into vacuum and material are equal and the neutrons are depolarized completely, then the fraction of incoherently scattered neutrons with a spin flip is 2/3. That is, the ratio of the spin flip probability and the probability of anomalous losses should be 2/3 for this model [11].

However, such a prediction does not have a sound theoretical basis; it comes from some preliminary work examining the possibility of UCN with energy lower than the critical energy of the substance becoming trapped [12, 13]. In the next section we present some theoretical calculations based on this work, which discuss the behavior of UCN scattering incoherently on material defects.

## 5. The capture cross-sections for UCN on material defects

The problem of anomalous losses of UCN in storage bottles led to the idea of localization of UCN [12]. In this section we give a simple theoretical description of neutron localization by analysis of neutron scattering on material defects.

The interaction of UCN with matter is described by the Schrödinger equation:

$$\left(\nabla^2 + \frac{2m}{\hbar^2} E_0\right)\psi(\mathbf{r}) = \frac{2m}{\hbar^2} V(\mathbf{r})\psi(\mathbf{r}). \qquad (1)$$

Following the standard development, the nuclear potential $V(\mathbf{r})$ may be approximated by a Fermi pseudo-potential,

$$V(\mathbf{r}) = \sum_i \frac{2\pi\hbar^2}{m} b_i \cdot \delta(\mathbf{r} - \mathbf{r}_i), \qquad (2)$$

where the constants $b_i$ have the sense of scattering lengths for bound nuclei. In the case of a homogeneous medium, this potential can be replaced by an average value, $V(\mathbf{r}) \approx U - iW$, where $U = \frac{2\pi\hbar^2}{m}\rho\,\mathrm{Re}\,b$, $W = \frac{2\pi\hbar^2}{m}\rho\,\mathrm{Im}\,b$ and $\rho$ is the nuclear density. As normal, the imaginary part of the potential describes absorption and inelastic scattering. The typical values are: $U \approx 10^{-7}$ eV, $|W/U| \approx 10^{-4}$.

For the case $E_0 > U$ the solutions of equation 1 are plane waves $e^{ikz}$ with the wave vector value $k = \frac{1}{\hbar}\sqrt{2m(E_0 - U + iW)} \equiv k' + ik''$, $k' \approx \frac{1}{\hbar}\sqrt{2m(E_0 - U)}$, $k'' \approx \frac{1}{2\hbar}\frac{\sqrt{2mW}}{\sqrt{E_0 - U}}$ and for the case $E_0 < U$ the solutions are damped waves $e^{-\ae z}$ with the wave vector value $\ae = \frac{1}{\hbar}\sqrt{2m(U - E_0 - iW)} \equiv \ae' - i\ae''$, $\ae' \approx \frac{1}{\hbar}\sqrt{2m(U - E_0)}$, $\ae'' \approx \frac{1}{2\hbar}\frac{\sqrt{2mW}}{\sqrt{U - E_0}}$.

The coherent reflection coefficient for UCN with energy $E_0$ from the material at normal incidence, $R$, is given by

$$R = \left|\frac{k_0 - k}{k_0 + k}\right|^2, \qquad (3)$$

where $k_0 = \frac{1}{\hbar}\sqrt{2mE_0}$ and $k = \frac{1}{\hbar}\sqrt{2m(E_0 - U + iW)}$.

That is, the loss factor for coherent reflection of UCN ($E_0 < U$) will be

$$\mu = 1 - R = 2\eta\sqrt{E_0/(U - E_0)}, \qquad (4)$$

where $\eta = \mathrm{Im}\,b / \mathrm{Re}\,b$.

Using the optical theorem, $\sigma_{tot.} = \frac{4\pi}{k_0}\mathrm{Im}\,b$, the relation (4) can be rewritten as

$$\mu = 2\sigma_{tot}.\rho\lambdabar\frac{E_0}{U}, \quad (5)$$

where $\lambdabar = \frac{\hbar}{\sqrt{2m(U-E_0)}}$ is the UCN penetration depth into the material. $E_0/U$ is the UCN probability amplitude at the surface. Equation 5 gives an expression for the reduction of UCN flux from crossing a surface layer of thickness $\lambdabar$.

For making further estimates we can use a simplified formula with $E_0$ set to $U/2$:

$$\mu = 2\eta = \sigma_{tot}.\rho\lambdabar \quad (6)$$

We now consider the influence of material defects on the process of UCN reflection. In crossing from the normal material into a defect, the material density will suddenly decrease. This means that incoherent scattering of UCN can occur at the boundary. In addition, the presence of hydrogen ($H_2O$, $H_2$) within the defects will also give incoherent scattering with depolarization. We will consider the simplest case, when the size of the defect (100 Å) is much less than the neutron wavelength (1000 Å) in which case the potential for the defect can be described by a $\delta$-function and solution of the Schrödinger equation (1) with the following form of potential is required:

$$V(\mathbf{r}) \approx U - iW + \frac{2\pi\hbar^2 \cdot B}{m}\delta(\mathbf{r}). \quad (7)$$

Here $B = nb$ is the scattering length for the defect, which is the scattering length of the nuclei in the defect multiplied by $n$, the number of nuclei inside the defect.

The Green's function for the case $E_0 > U$ may be written as

$$G(\mathbf{r},\mathbf{r}') = -\frac{1}{4\pi}\frac{e^{ik|\mathbf{r}-\mathbf{r}'|}}{|\mathbf{r}-\mathbf{r}'|}, \quad (8)$$

where $k$ is the absolute value of the neutron wave vector and when $E_0 < U$ (the most interesting case)

$$G(\mathbf{r},\mathbf{r}') = -\frac{1}{4\pi}\frac{e^{-\ae|\mathbf{r}-\mathbf{r}'|}}{|\mathbf{r}-\mathbf{r}'|} \quad (9)$$

with æ as defined above and where, in this instance, æ″ describes the flux of neutrons below the potential barrier.

In the Born approximation, the solution of the Schrödinger equation (1) with the potential (7) for neutrons with energy above the energy barrier $E_0 > U$ is

$$\psi = e^{i\mathbf{k}\mathbf{r}} + F \cdot \frac{e^{ikr}}{r}, \quad (10)$$

where $F = -B$ is the scattering amplitude and below the barrier ($E_0 < U$), the wave function may be written as:

$$\psi = e^{\ae z_0}\left(e^{-\ae z} + F \cdot \frac{e^{-\ae r}}{r}\right), \quad (11)$$

where $z$ is the normal to the surface and $z_0$ is the value of the coordinate fixing the plane of surface (We need to introduce the finite distance from the surface because of the damped character of the solution).

Using the general expression for the probability density flux,

$$\mathbf{j} = \frac{i\hbar}{2m}\left(\psi\,\vec{\nabla}\psi^* - \psi^*\,\vec{\nabla}\psi\right) \quad (12)$$

and the wave functions (equations 10 and 11), we obtain a value for the incident flux density in the medium:

$$j_{incident} = \frac{\hbar k'}{m} e^{-2k''z}. \quad (13)$$

The absorption cross-section is given by the ratio of the net flux through a sphere of radius $r$ surrounding the center of defect to the magnitude of the incident flux (equation 13), taken with a negative sign:

$$\sigma_{capt.} = -\frac{\int_S j r^2 d\Omega}{j_{incident}}. \quad (14)$$

The flux $j$ in equation 14 is the radial part of the flux calculated from equation 12 with the wave function equation 10.

As $r \ll \lambda$, only the $s$-waves give contributions to $\sigma_{capt.}$, so for $j$ in equation 14 we can use an $s$-wave expansion of equation 10,

$$\psi \approx \frac{e^{ikr}}{r}\left(F + \frac{1}{2ik}\right) - \frac{1}{2ik}\frac{e^{-ikr}}{r}. \quad (15)$$

The flux, $j$, calculated using the function (15) can be represented by the sum of three terms:

- the contribution of the incoming wave

$$j_{in} = -\frac{\hbar}{m}\frac{k'}{4|k|^2}\frac{e^{2k''r}}{r^2}, \quad (16)$$

- the contribution of the outgoing wave

$$j_{out} = \frac{\hbar k'}{m}\frac{e^{-2k''r}}{r^2}\left|F + \frac{1}{2ik}\right|^2, \quad (17)$$

- the contribution arising from the interference of the incoming and outgoing waves

$$j_{interf.} = \frac{\hbar}{m}k''\,\text{Re}\left\{\frac{e^{2ik'r}}{r^2}\left(F + \frac{1}{2ik}\right)\frac{1}{k^*}\right\}, \quad (18)$$

where $k^* = (k' + ik'')^* = k' - ik''$.

Using the sum of the fluxes $j_{in}$, $j_{out}$ and $j_{interf.}$ in the integral (14) and taking the limit of $r$ equal to zero, one obtains the following formula for the absorption cross-section

$$\sigma_{capt.} = 4\pi \left( \frac{\operatorname{Im} F}{k'} - |F|^2 \right), \tag{19}$$

where $4\pi |F|^2$ is the elastic scattering cross-section and

$$\sigma_{tot.} = 4\pi \frac{\operatorname{Im} F}{k'} \tag{19'}$$

is the analog of the optical theorem for the material defect.

For discussion of anomalous reflection coefficients, it is relevant to study the behavior of the absorption cross-section for the case $E_0 < U$. The cross-section values in this case come straight from equations 19 and 19' with replacement of $k$ by $i\ae$ and $k'$ by $\ae''$. The absorption and total cross-sections are given by:

$$\sigma_{capt.}\big|_{E_0<U} = 4\pi \left( \frac{\operatorname{Im} F}{\ae''} - |F|^2 \right). \tag{20}$$

$$\sigma_{tot.def.}\big|_{E_0<U} = \frac{4\pi \operatorname{Im} F}{\ae''} \tag{20'}$$

Relating the size of the absorption cross-sections for the two cases: $\sigma_{capt.}\big|_{E_0<U} = \frac{|U-E_0|}{W} \sigma_{capt.}\big|_{E_0>U}$, and remembering that $k'/\ae'' = \frac{E_0-U}{W} \approx 10^4$, we see that the magnitude of the absorption cross-section in the case $E_0 < U$ is about 10'000 times larger than in the case $E_0 > U$. This enhancement of the capture cross-section arises from localization of the UCN with $E_0 < U$ around the material defect in accordance with wave function (9).

## 6. Estimation of losses in reflections from Beryllium

Equations 20 and 20' give a new mechanisms, a fairly strong incoherent component to scattering from material defects, that needs taking into account when estimating the UCN loss factor under reflection. We present estimates for the reflection losses by Beryllium from using formula 20' for the total cross-section ($\sigma_{tot.def.}$) for UCN scattering on material defects and the simplified equation:

$$\mu_{an} = N_{def.} \sigma_{tot.def.} \lambdabar, \tag{21}$$

where $N_{def.}$ is the defect density.

The density of defects in Beryllium can be estimated from the measured macroscopic cross-sections for very cold neutrons (the macroscopic cross-section at 30 K is dominated by scattering on the defects and of the order 1 to 2 cm$^{-1}$, see Fig. 3) and values for the size and density of the defects. Assuming a defect size $L = 100$ Å, a defect volume of about $(100 \text{ Å})^3$ and a material density inside the defect of about 50%, then the scattering cross-section on defects containing $n$ atoms of Beryllium ($n = 0.5 \rho (100 \text{ Å})^3 = 0.6 \cdot 10^5$) with a scattering length ($b = 0.8 \cdot 10^{-12}$ cm) will be:

$$\sigma_{scatt.def.} = 4\pi (nb)^2 = 2.9 \cdot 10^{-14} \text{ cm}^2, \tag{22}$$

which gives a value for the defect density $N_{def.} = = 3.5$ to $7 \cdot 10^{13}$ cm$^{-3}$.

Using the expression for æ″ and for $W$, the total cross-section for UCN interactions with the defects in the energy region $E_0 < U$ is

$$\sigma_{tot.def.} = \frac{4\pi \operatorname{Im}(nb)}{æ″} = \frac{2n}{\lambdabar \rho} = 0.8 \cdot 10^{-12} \text{ cm}^2. \qquad (23)$$

Noting that half of the UCN scattered on defects return to the vacuum and using equation 23 for $\sigma_{tot.def.}$, we obtain:

$$\mu_a = N_{def.} \frac{\sigma_{tot.def.}}{2} \lambdabar = \frac{N_{def.} n}{\rho} = (1.7 \div 3.4) \cdot 10^{-5} \qquad (24)$$

and the result that the losses from scattering on material defects are proportional to the ratio of the incoherent scattering atoms to the total number of atoms. This result is rather obvious.

## 7. Discussion

The reflection coefficient for Beryllium taking into account the effect of scattering of UCN by material defects (1.7 to 3.4 $\cdot 10^{-5}$), agrees rather well with the experimental values ($3 \cdot 10^{-5}$), although it should be remarked that the values used for the size of the defects and the density of atoms inside the defects are, in the first instance, somewhat arbitrary. Also, the energy dependence of UCN losses on reflection by material defects is different from that predicted by neutron optics calculations. Equation 5 shows that the energy dependence of UCN normal losses is proportional to $\sqrt{E_0}/\sqrt{U - E_0}$, but the energy dependence of the UCN losses on material defects is proportional $E_0$; this comes from the different energy dependence of the capture cross-section for atoms and defects. The dashed line in Fig. 2 shows the fit of the experimental results to the expected energy dependence, $\mu_{an} \sim E_0$: It is in good agreement with the measured values.

The agreement between the model and measurement for reflection from Beryllium can be considered as a strong indication that the typical size of defects has to be about 100Å in order to provide the amount of incoherent scattering necessary. This is about the $\lambdabar$-depth for UCN penetration into material (100-150 Å). In fact, equation 23 shows that total cross-section for UCN interactions with material defects $(\sigma_{tot.def.})$ is proportional to $n$ or $L^3$. This means that $\sigma_{tot.def.}$ increases very fast with size of the defect when $L < \lambdabar$. The total cross-section is equal to the cross-section of the defect then $L \geq \lambdabar$.

The theoretical development in section 5 is quite general and high reflection losses will be observed with any crystalline substance with a defective structure. For example, the macroscopic cross-section for elastic scattering on the defects of Aluminum and Zirconium foils is of the order 10 to 20 cm$^{-1}$. These values for the macroscopic cross-section suggest a considerably higher defect density than for Beryllium and that UCN reflection coefficients might be about 3 to $6 \cdot 10^{-4}$.

The presence of hydrogen (in the form of water molecules or gas atoms/molecules) in the defects leads to enhanced up-scattering and capture of UCN by the hydrogen. The losses for this case will be given by a modified version of equation 24:

$$\mu_a = \frac{N_{def} \cdot n_n}{\rho} \cdot \frac{\operatorname{Im} b_n}{\operatorname{Im} b_m}, \qquad (25)$$

where $n_n$ is the numbers of hydrogen atoms in the defect, $b_n$ is average scattering length for hydrogen atoms and $b_m$ is average scattering length of material. Thus, the experimentally observed up-scattering of UCN [3], where the amount of hydrogen present was much less than allowed explanation in the context of neutron optics calculations, can be explained by the enhanced up-scattering of UCN by Hydrogen inside defects.

The presence of hydrogen atoms in the inter-crystalline defect explains the effect of UCN depolarization at storage.

Perfect mono-crystals and the amorphous substances with uniform density can be considered as materials without anomalous UCN reflection losses. Probably, the frozen Fomblin oil at a temperature not far below the freezing point is just an example of a substance with a very low level of defects. Therefore anomalous losses were not observed for the low temperature Fomblin oil and the UCN losses factor of $2 \cdot 10^{-6}$ is the minimum among the studied materials.

The work has been supported by the Russian Foundation for Basic Research (Grant No 04-02-17440).

The authors are grateful to F. Atchison for remarks and editing of the manuscript. We are grateful to S. Ivanov, V. Plakhty and Yu. Pokotilovski for useful critical remarks on the text of the article.

**References**

1. L.V.Groshev, V.N.Dvoretsky, A.M.Demidov et al., Phys. Lett. **B 34** (1971) 293.
2. V.M.Lobashev, G.D.Porsev, A.P.Serebrov, Preprint LNPI –37, Gatchina, (1973); A.I.Egorov, V.M.Lobashev, V.A.Nazarenko et al., Yad. Fiz. **19** (1974) 300, Sov. J. Nucl. Phys. **19** (1974) 147.
3. A.V. Strelkov, M.Hetzelt, Zh. Eksp. Teor. Fiz. **74** (1978) 23; Sov. Phys. JETP **47** (1978) 11.
4. J.C.Bates, Phys. Lett. **A 88** (1982) 427.
5. P.Ageron, W.Mampe, A.I.Kilvington, Z. Phys. B – Condensed Matter **59** (1985) 261.
6. V.Nesvizhevskii, A.Serebrov, R.Tal'daev et al., Sov. Phys. JETP **75** (3) 1992.
7. A.Serebrov, V.Varlamov, A.Kharitonov et al., Preprint PNPI–2564, Gatchina, (2004), p. 25.
8. A.Steyerl, B.Yerozolimsky, A.Serebrov et al., Eur. Phys. J. **B 28** (2002) 299–304.
9. Yu.Pokotilovski JETP Vol. **96**, No 2, 2003.
10. A.Serebrov, A.Vasiliev, M.Lasakov et al. NIMA **440** (2000) 717–721.
11. A.Serebrov, M.Lasakov, A.Vasiliev et al. Phys. Lett. **A 313** (2003) 373–379.
12. A.Serebrov. Preprint PNPI–2193, Gatchina, (1997), p. 7.
13. A.Serebrov, N.Romanenko. Preprint PNPI–2194, Gatchina, (1997), p. 8.